\documentclass[aps,prd,preprint,groupedaddress]{revtex4-1}
\usepackage{amssymb,amsmath}
\usepackage{graphicx} 
\usepackage{dcolumn}  
\usepackage{epsfig}   
\usepackage{hyperref}
\usepackage{ifpdf}
\usepackage{float}
\usepackage{longtable}

\newsavebox{\accentbox}
\begin{document}

\preprint{}

\title{A PHYSICAL VERSION OF THE QCD CONFINEMENT SCALE(S)}



\author{H. M. Fried$^{\dag}$, P. H. Tsang$^{\dag}$}

\affiliation{${}^{\dag}$ {Physics Department, Brown University, Providence, RI 02912, USA} }

\date{\today}

\begin{abstract}
We suggest a physical definition of the confinement mass scale in QCD in the framework of non-perturbative, gauge invariant QCD, where all possible gluons exchanged between any pair of quark lines are included; and we insist that a stable, quark bound state should not and must not have transverse quark fluctuations larger than the Compton wavelength of the bound state particle itself. This is possible in our QCD formulation because there are two parameters which describe confinement, a mass scale $\mu$, and a "deformation parameter" $\xi$, which shrinks the transverse-quark-coordinate separation distribution $\varphi(b)$ away from Gaussian. With the mass scale $\mu$ defined as equal to the mass of each quark bound state, we show that $\xi$ decreases with increasing bound state mass, $m_{BS}$, using order-of-magnitude estimates which agree with obvious intuition. Our $\xi$-values, including a calculation for the recently detected 4-quark system, display the predicted behavior: $\xi$ decreases with increasing $m_{BS}$. Our results for $\varphi(b)$, when the quark bound state is a nucleon or heavier, then show agreement with the form of Gaussian momentum-space fall-offs in recent Light-Front holographic analyses.
\end{abstract}


\maketitle

\section{\label{SEC1}Introduction}

The purpose of these remarks is to suggest a physical definition of the confinement mass scale in QCD, associated with the non-perturbative binding of quarks into hadrons by the gauge-invariant exchange of all possible gluons between all relevant quarks, with cubic and quartic gluon interactions included \cite{FriedGabellini2010,2,3,4,5,6}. For simplicity and clarity, the hadrons considered at first are "qualitative" in the sense that only one flavor of quark is considered, in the absence of weak and electromagnetic effects, simplifications which can easily be rectified; these hadrons will be denoted as pions and nucleons, plus possible 4-quark states, and any possible future bound state of multi-quark systems. Correct quark flavors and masses are used for a final computation.

The confinement scales defined here will turn out to be Different; but yet they are all the Same. They are Different in that each stable bound state has a different mass scale, but they are the Same in that the confinement mass scale always equals the mass of that bound state. This has a simple physical basis in that the transverse fluctuations of the bound quarks which comprise the bound state particle cannot and should not be larger than the Compton wavelength of that particle, thereby preventing a plethora of unwanted interactions completely at variance with the experimentally observed properties of a hadron. A "nucleon" whose electrically-charged quarks can naturally separate to distances much larger than $10^{-13}cm$. is not a true nucleon but rather an unstable object waiting to be split into its constituent quarks under atomic or external electromagnetic fields.

In order for this scheme to work there must be an additional parameter other than the mass scale, which enters into the confinement analysis, and whose numerical value is able to change, if only slightly, in the description of higher-mass bound states. This is the "deformation parameter" $\xi$ of Ref.~\cite{2}, wherein the probability of transverse-quark-coordinate separation is given by a normalized distribution

\begin{equation*}
\varphi(b)=\varphi(0)\ e^{\displaystyle{-(\mu b)^{2+\xi}}} ,
\end{equation*}

\noindent where for a pion, $b$ is the transverse separation of a $q$ - $\bar{q}$ pair forming (the major component of) a pion, $\mu$ is the mass scale such that transverse fluctuations larger than $\mu^{-1}$ give little contribution, and $\xi$ is a small, real, positive parameter on the order of $0.1$ for this bound state. The corresponding potential $V(r)$ binding this $q$ - $\bar{q}$ pair was there derived to be \begin{equation*}
V(r) \simeq \xi \mu (\mu r)^{{1+\xi}},
\end{equation*}
\noindent and in the pion analysis of Ref.~\cite{2}, $\mu$ was taken to be on the order of the pion mass, $m_\pi$. The change in viewpoint expressed in this paper is that, for this pion calculation, $\mu =m_\pi$, and that $\mu$ will be defined as the bound state mass of any higher-mass particle, with a correspondingly small decrease in the value of $\xi$.

There is a very simple, intuitive reason why one might expect a smaller $\xi$-value to appear for heavier quark bound states. The probability function $\varphi(b)$ is an essential part of the eikonal function $\Xi(b)$, of the high-energy scattering amplitude of a $q$ and $\bar{q}$; and with a well-defined relation ~\cite{2} between $\Xi(b)$ and effective scattering/binding potentials, it is easy to obtain $V(r)$ from $\Xi(b)$ with no extra assumption of a static $q$-$\bar{q}$ situation, an assumption that is wrong in principle and in practice ~\cite{FriedGabellini2010,2,3,4,5,6}.

With this procedure it is instructive to ask what is the result for $V(r)$ found for a perfect Gaussian, with $\xi = 0$. Immediately, that calculation produces a zero result because a Gaussian is "too symmetric", and there is no reason to expect smaller, rather than larger values of $b$ to be enhanced. However, for $\xi>0$ such "enhancement" is intuitively clear, because $\varphi(b)$ is "bunched" or "enhanced" for smaller values of $b$, since $\varphi(b)$ now falls off more rapidly than a Gaussian.

For the 3-quark problem, what value of $\xi$ would be appropriate, and -- intuitively -- would it be larger or smaller than that of the pion calculation, $\xi \sim 0.1$? Here there are 3 quarks, with each of them exchanging a Gluon Bundle with two other quarks; and a smaller value of $\xi$ should be needed, since each quark is being pulled by two others. One finds, below, that the qualitative solutions for $\xi$ produce just this behavior, $\xi \sim 0.01$, when the relevant scale factor $\mu$ is required to equal the bound state, nucleon mass.

For quark binding into a nucleon, the same $V(r_{ij})$ is the potential acting between any two quarks whose instantaneous transverse separation is $|\vec{b}_i - \vec{b}_j|$, and whose 3-dimensional separation is $r_{ij} = |\vec{r}_i - \vec{r}_j  |$. In the following, and based on the pion analysis, the flux tube or Gluon Bundle coordinates which interact between any two quarks are assigned the order-of-magnitude value $r_{ij} = (m_\pi)^{-1}$, while the mass scale $\mu$ is chosen to be that of the nucleon. The corresponding $\xi$ value will then decrease, and this has an interesting effect when fitting p-p elastic scattering data, which depends upon the Fourier transform of $\varphi(b)$, becoming for small $\xi$ very close to the Gaussian $\exp(-\vec{q}^{\ 2} / 4m_{p}^2)$, where $m_p$ is the proton mass, which is the form of momentum space fall-off suggested in recent light-front holographic analyses~\cite{sjb}. A prior calculation, in the context of the work of Ref.~\cite{sjb}, also required the maximum, effective quark transverse fluctuations to be less than $10^{-13}\ cm.$~\cite{sjb08}. 

The formalism of Refs.~\cite{FriedGabellini2010,2,3,4,5,6} provides a simple way to insure that bound quark fluctuations are always less than the Compton wavelength of the bound state particle, by arguing that the confinement mass scale for each bound state should be understood to be the mass of that bound state. This perhaps novel idea can easily be employed with the restriction that the small $\xi$-parameter can be decreased for larger values of the bound state mass, until a point is reached where it is so small as to be negligible. For example, in the construction of a model deuteron of Ref.~\cite{3}, $\xi$ was quite irrelevant to the 2.2 MeV bound state of the deuteron, and was simply neglected. Better approximations to high energy pp elastic scattering, now underway, can easily retain all $\xi$-dependence associated with the proton mass of the scattering bound states; but again, that dependence associated with the numerical value of $\xi$ is expected to be insignificant.

\section{\label{SEC2}Formulation}
In our formulation in Ref.~\cite{2}, the confinement mass parameter $\mu$ first appears in the transverse-fluctuation probability function $\varphi(b)$, so that $b$ values larger than $1/\mu$ give a negligible contribution to any Fourier transform of powers of $\varphi(b)$. The procedure adopted here, in contrast to that of Ref.~\cite{2}, is to define a simpler, approximate equation for the bound-state-energy, or the mass $m_{BS}$ of the hadron so defined, but whose Order-of-Magnitude (OoM) solutions easily display a decreasing OoM of the parameter $\xi$ with increasing $m_{BS}$.

This new, approximate relation is obtained in three steps: i) Neglect all kinetic energy contributions of the $qs$ and/or $\bar{q}s$, since in such small confined spaces all such $KE$ contributions must be non-relativistic; ii) underestimate the contributions of each $q/\bar{q}$ $V(r_{ij})$ interaction by replacing its $r_{ij}$ by $1/m_{\pi}$, since that replacement is the essence of the OoM of the pion calculation; iii) and replace $\mu$ by the $m_{BS}$: 
\begin{equation}\label{eq1}
E_0 \rightarrow m_{BS} \simeq n_q m_q + \xi \sum_q m_{BS}\bigg(\frac{m_{BS}}{m_{\pi}}\bigg)^{1+\xi},
\end{equation} where $\sum_q$ represents the number of pairwise $q$ and/or $\bar{q}$ interactions. One them obtains 

\begin{equation}\label{eq2}
m_{BS} \simeq n_q m_q + \frac{n_q(n_q -1)}{2} m_{BS}\ \xi \bigg(\frac{m_{BS}}{m_{\pi}}\bigg)^{1+\xi};
\end{equation} and since $\xi$ is expected to be $\ll 1$, ~\eqref{eq2} simplifies further to 
\begin{equation}\label{eq3}
\xi \simeq [1-\frac{n_qm_q}{m_{BS}}]\cdot \bigg(\frac{1}{\frac{n_q(n_q-1)}{2}(\frac{m_{BS}}{m_{\pi}})}\bigg)
\end{equation}

Note that, if the $m_{BS}$ of ~\eqref{eq3} is itself chosen to be $m_{\pi}$, the value of $\xi$ will turn out to be an OoM larger than that of Ref.~\cite{2}, $\xi \approx 1.0$, instead of $0.1$. But for larger values of $m_{BS}$, where the ratio $(\frac{m_{BS}}{m_{\pi}})$ of ~\eqref{eq3} makes a difference, the $\xi$-values found are reasonable. For example, for a proton, ~\eqref{eq3} yields $0.05$, suggesting that the result of using $\xi=0$ when incorporating Fourier transforms into data-fitting prescriptions would be barely distinguishable from using a more correct value of $\xi$. 

For 4-quark ~\cite{9} and higher bound states, their value of $\xi$ would be even smaller, using proper masses for the $c$, $u$, and $d$ quarks. There then would be little difference between the Fourier tansform of $\varphi(b)$ and a pure Gaussian, but a Gaussian which involves the mass of that bound state.  Table~\ref{table:xivalues} displays the expected variations of $\xi$ with respect to $x$, where $x=\frac{m_{BS}}{m_{\pi}}$, using correct quark flavor masses for known hadrons.



\begin{longtable}{|l|l|l|l|l|}
\hline
Hadron      & $m_{BS}$ [$GeV/c^2$] & Quark Content & x &  $\xi$  \\
\hline
Proton                        & 0.938    & uud     & 6.7     & 0.05              \\
Neutron                       & 0.939    & udd     & 6.7      & 0.05     \\
Lambda,$\Lambda^0$            & 1.116   & uds     & 8.0     & 0.04     \\
Charmed Lambda, $\Lambda_c^+$ & 2.286  & udc     & 16.3     & 0.009   \\
bottom Lambda, $\Lambda_b^0$  & 5.619   & udb     & 40.1     & 0.008   \\
Sigma+, $\Sigma^+$ & 1.189  & uus   & 8.5         & 0.03            \\
Sigma0, $\Sigma^0$ & 1.193  & uds   & 8.5         & 0.03     \\
Sigma-, $\Sigma^-$ & 1.197 & dds   & 8.6          & 0.03    \\
Charmed Sigma, $\Sigma^{++}_c$ & 2.454  & uuc      & 17.5     & 0.009   \\
Bottom Sigma, $\Sigma_b^+$   & 5.811    & uub       & 41.5     & 0.008     \\
Bottom Sigma, $\Sigma_b^-$     & 5.816   & ddb     & 41.5      & 0.008  \\
Xi, $\Xi^0$ & 1.315   & uss   &  9.4       & 0.03     \\
Xi, $\Xi^-$          & 1.322   & dss      & 9.4     & 0.03     \\
Charmed Xi, $\Xi_c^+$ & 2.468   & usc     &  17.6     & 0.008    \\
charmed Xi, $\Xi_c^0$  & 2.471  & dsc    &  17.6      & 0.008    \\
charmed Xi prime, ${\Xi'}_c^+$   & 2.575    & usc  & 18.4        & 0.008    \\
charmed Xi prime, ${\Xi'}_c^0$   & 2.578    & dsc  & 18.4         & 0.008    \\
double charmed Xi, $\Xi_{cc}^+$    & 3.519    & dcc & 25.1          & 0.004   \\
bottom Xi (Cascade B), $\Xi_b^0$ & 5.788    & usb  & 41.3         & 0.008    \\
bottom Xi(Cascade B), $\Xi_b^-$ & 5.791    & dsb   &  41.4       & 0.008     \\
charmed Omega, $\Omega_c^0$  & 2.695    & ssc    & 19.3      & 0.008            \\
bottom Omega, $\Omega_b^-$ & 6.071    & ssb      &  43.4    & 0.007 \\
\hline
\caption{$\xi$-values for the known hadrons, using correct quark flavor masses~\cite{particles}, displays the expected variations in $\xi$ as a function of $x$ where $x=m_{BS}/m_\pi$}
\label{table:xivalues}
\end{longtable}

\section{\label{SEC7}SUMMARY}

In summary, the proposal made in the above paragraphs, in the context of explicit quark transverse fluctuations, allows one to be certain that such fluctuations will not exceed the Compton wavelength of the bound state formed from such quarks and/or antiquarks. This is surely a physical requirement; and it can be seen, at least qualitatively, on the basis of the non-perturbative, gauge-invariant, functional formulation of Ref.~\cite{2}, in which the impossibility of measuring transverse quark components was explicitly built into the quark-gluon interaction Lagrangian.

\begin{acknowledgments}
This publication was made possible through the support of a Grant from the Julian Schwinger Foundation. The opinions expressed in this publication are those of the authors and do not necessarily reflect the views of the Julian Schwinger Foundation. We especially wish to thank Mario Gattobigio for his many kind and informative conversations relevant to the Nuclear Physics aspects of our work. It is also a pleasure to thank Mark Restollan, of the American University of Paris, for his kind assistance in arranging sites for our collaborative research when in Paris. 
\end{acknowledgments}

\end{document}